\documentclass[12pt]{article}

\usepackage{float}
\usepackage{graphics, graphicx}
\usepackage{amsfonts}
\usepackage{ natbib}

\usepackage[margin=10pt,font=small,labelfont=bf,labelsep=period]{caption}

\usepackage{amsmath, amssymb, amsthm, graphicx, subfigure}

\newtheorem{thm}{Theorem}
\newtheorem*{rem}{Remark}
\newtheorem{lem}{Lemma}

\newcommand{\ben}{\begin{enumerate}}
\newcommand{\een}{\end{enumerate}}
\newcommand{\beq}{\begin{equation}}
\newcommand{\eeq}{\end{equation}}
\newcommand{\beqn}{\begin{eqnarray}}
\newcommand{\eeqn}{\end{eqnarray}}
\newcommand{\bea}{\begin{eqnarray*}}
\newcommand{\eea}{\end{eqnarray*}}
\newcommand{\bs}{\boldsymbol}

\newcommand{\E}{\mathop{\mathbb E}}
\newcommand{\RR}{\mathop{\mathbb R}}

\newcommand{\mysup}{\operatornamewithlimits{sup}}

\newcommand{\asym}{\stackrel{\mathrm{asym}}{\sim}}

\newcommand{\dlaw}{\stackrel{\mathrm{\mathcal{L}}}{\longrightarrow}}

\newcounter{Aequ}
\newenvironment{AEquation}
  {\stepcounter{Aequ}%
    \addtocounter{equation}{-1}%
    \renewcommand\theequation{A.\arabic{Aequ}}\equation}
  {\endequation}

\newcounter{Aaux}
\newenvironment{AAlign}
  {\setcounter{Aaux}{\theequation}
    \setcounter{equation}{\theAequ}%
    \renewcommand\theequation{A.\arabic{equation}}
    \align}
  {\endalign\setcounter{Aequ}{\value{equation}}\setcounter{equation}{\theAaux}}

\begin{document}

\title{Statistical testing for conditional copulas}

\author{
{\bf Elif F. Acar}\\
Department of Mathematics and Statistics, McGill University\\
{\bf Radu V. Craiu}\\
Department of Statistics, University of Toronto\\
and\\
{\bf Fang Yao }\\
Department of Statistics, University of Toronto}

\date{}

\maketitle

\begin{abstract}
In conditional copula models, the copula parameter is deterministically linked to a covariate via the calibration function. 
The latter is of central interest for inference and is usually estimated nonparametrically. However, when a parametric model for the calibration function is appropriate, the resulting estimator exhibits significant gains in statistical efficiency and requires smaller computational costs. We develop methodology for testing a parametric formulation of  the calibration function against a general alternative and propose a generalized likelihood ratio-type test that enables conditional copula model diagnostics.
We derive the asymptotic null distribution of the proposed test and study its finite sample performance using simulations. The method is applied to two data examples.
\end{abstract}


\newpage
\section{Introduction}
\label{s:sec1}

Copulas are an important tool for modeling dependence. The recent development of conditional copulas by \cite{Patton:2006} widely expands the range of possible applications, as it allows covariate adjustment in copula structures and thus enables their use in regression settings. 
Specifically, if  $X$ is a  covariate that affects the dependence between the continuous random variables $Y_1$ and $Y_2$, then the conditional joint distribution $H_x$ of $Y_1$ and $Y_2$ given $X=x$ can be written as
$
H_x (y_1, y_2 \mid x) = C_x\{F_{1|x}(y_1 \mid  x), F_{2|x}(y_2 \mid  x) \mid x \},
$
where $F_{i |x}$ is the conditional marginal distribution of $Y_i$ given $X=x$ for $i=1,2$ and $C_x$ is the conditional copula, i.e. the joint distribution of $U_1 \equiv F_{1x}(Y_{1} \mid x)$ and $U_2 \equiv F_{2x}(Y_{2} \mid x)$ given $X=x$.

When the dependence structure is in the inferential focus, one needs to specify a functional model between the covariate  $X$ and the copula $C_x$. 
In the context of a parametric copula family, \cite{Acar/Craiu/Yao:2011} have studied a nonparametric estimator of the calibration function $\eta(X)$ in
\beq
\label{eq:1}
(U_{1}, U_{2}) \mid X=x \sim C_x\{u_1, u_2 \mid \theta(x)= g^{-1}(\eta(x))\},
\eeq
where $g: \Theta \rightarrow \mathbf{R}$ is a known link function that allows unrestricted estimation for $\eta$. 

It is known that if a parametric model for $\eta(X)$ is suitable, then fitting a nonparametric model leads to an unnecessary loss of efficiency. 
For instance, in Table 1 in \cite{Acar/Craiu/Yao:2011} this loss is illustrated in the case of an underlying linear calibration function.
Furthermore, parametric formulation of $\eta(X)$ yields a much simpler conditional copula model that is more convenient for subsequent analysis.
Therefore, it is of great practical importance to determine whether $\eta(X)$ can be reasonably estimated using a simple parametric form. 
While one can construct pointwise confidence intervals as in \cite{Acar/Craiu/Yao:2011} and check whether an estimated parametric calibration function falls within the confidence intervals, such visual inspections are not sufficient to make valid inference on the form of the calibration function. 
One needs to construct simultaneous confidence intervals across the covariate range or rigorous hypothesis tests for the specification of the calibration function. Here we take the latter approach.

Our development focuses on the hypotheses of the form $\rm{H}_0:  \mbox{``}\eta(\cdot)  \mbox{ is linear in $X$"} $ versus $\rm{H}_1: \mbox{``}\eta(\cdot) \mbox{ is not linear in $X$" }$ under the conditional copula model in (\ref{eq:1}).
This class of hypotheses includes the important special case of $\rm{H}_0:  \mbox{``}\eta(\cdot)  \mbox{ is constant" }$ versus $\rm{H}_1: \mbox{``}\eta(\cdot) \mbox{ is not constant"}$. This is of particular interest because in those cases where $\eta$ can be reasonably estimated by a constant, one can rely on statistical methods developed for the classical copula model.

However, such hypotheses cannot be tested using the canonical likelihood ratio test  (LRT) because estimation under the alternative hypothesis is performed nonparametrically. Exploration of the asymptotic distribution of the ratio test falls within the scope of the generalized likelihood ratio test (GLRT) developed by \cite{Fan/Zhang/Zhang:2001} for testing a parametric null hypothesis versus a nonparametric alternative hypothesis.  Since nonparametric maximum likelihood estimators are difficult to obtain and may not even exist, \cite{Fan/Zhang/Zhang:2001} suggested using any reasonable nonparametric estimator under the alternative model. 
In particular, using a local polynomial estimator to specify the alternative model of a number of hypothesis testing problems, \cite{Fan/Zhang/Zhang:2001} showed that the null distribution of the GLRT statistic follows asymptotically a chi-square distribution with the number of degrees of freedom independent of the nuisance parameters.  
This result, referred to as \emph{Wilks phenomenon}, holds for Gaussian white-noise model \citep{Fan/Zhang/Zhang:2001}, varying-coefficient models, which include the regression model as a special case \citep{Fan/Zhang/Zhang:2001}, spectral density \citep{Fan/Zhang:2004b}, additive models \citep{Fan/Jiang:2005} and single-index models \citep{Zhang/Huang/Lv:2010}.

We expand the  GLRT-based approach to testing the calibration function in conditional copula models. 
The test procedure employs the nonparametric estimator proposed by \cite{Acar/Craiu/Yao:2011} when evaluating the local likelihood under the alternative hypothesis. The major contribution is the construction of a rigorous framework for such GLRT-based tests in the conditional copula context, which leads to improved efficiency when a suitable parametric form can be specified.
 It is worth mentioning that the proposal can easily accommodate the test for an arbitrary parametric form. The description of the test, the derivation of its asymptotic null distribution and the discussion of practical implementation are included in Section \ref{s:sec2}. The finite sample performance of the test is illustrated using simulations and two data examples in Section \ref{s:sec3} and  \ref{s:sec4}, respectively. The paper ends with concluding remarks.

\section{Generalized likelihood ratio test for copula functions} 
\label{s:sec2}

Suppose that $\{ (U_{11},U_{21}, X_1), \ldots, (U_{1n},U_{2n}, X_n)\} $ is a random sample from the conditional copula model \eqref{eq:1}. The hypothesis of interest is 
\beqn
\label{eq:2}
\rm{H}_0 : \eta(\cdot) \in \mathfrak{f}_L   \qquad \quad  \text{versus} \qquad \quad \rm{H}_1 : \eta(\cdot) \notin \mathfrak{f}_L ,
\eeqn
where $\mathfrak{f}_L= \{ \eta(\cdot): \exists \ a_0, a_1 \in \mathbb{R} \mbox{ such that } \eta(X)=a_0 + a_1 X,  \; \; \forall X\in \mathcal{X}  \}$ denotes the set of all linear functions on $\mathcal{X}$. 

In what follows, we assume that the density $c_x$ of $C_x$ exists and for simplicity we use the notation $\ell (t, u_1,u_2)  = \ln c_{x} \{ u_1 , u_2; g^{-1}(t) \}$. Furthermore, the first and second partial derivatives of $\ell$ with respect to $t$ are assumed to exist and are denoted by $\ell_{j} (t, u_1, u_2) = \partial^{j}  \ell (t, u_1,u_2) / \partial t^j $, for $j=1,2$.

\subsection{Proposed GLRT for the conditional copula model}
A natural way to approach \eqref{eq:2} is through the likelihood ratio of the {\em restricted} (i.e., conditional copula with a linear calibration function)
and the {\em full} (i.e., conditional copula with an arbitrary calibration function)
models, or equivalently, through the difference 
$$
 \sup_{\eta(\cdot) \notin \mathfrak{f}_L } \{\mathbb{L}_n (\rm{H}_1) \}  - \sup_{\eta(\cdot) \in \mathfrak{f}_L }  \{\mathbb{L}_n (\rm{H}_0)\},
$$
where 
\bea
\mathbb{L}_n (\rm{H}_0)&=& 
\sum_{i=1}^n  \ell ( a_0 +a_1 X_i, U_{1i}, U_{2i} ), \\
\qquad \quad \mathbb{L}_n (\rm{H}_1)&=& 
\sum_{i=1}^n  \ell ( {\eta}(X_i), U_{1i}, U_{2i} ).
\eea

The supremum of the log-likelihood function under the null hypothesis is given by
\bea
\mathbb{L}_n (\rm{H}_0, \tilde{\eta})= \sum_{i=1}^n \ell (  \tilde{\eta}(X_i),   U_{1i}, U_{2i} ).
\eea
where $\tilde \eta (X)= \tilde a_0 + \tilde a_1 X $, with $\boldsymbol{\tilde a} = ( \tilde a_0 ,  \tilde a_1)$ denoting the maximum likelihood estimator of the parameter $\boldsymbol{a} = (a_{0}, a_{1})$.
Under the alternative,  the general unknown form of $\eta(\cdot)$ adds significant  complexity to the calculation of  the supremum.  
We use the nonparametric estimator of $\eta(\cdot)$ proposed by \cite{Acar/Craiu/Yao:2011} to define the log-likelihood under the full model. 
Specifically, for each observation $X_{i}$ in a neighbourhood of an interior point $x$, we approximate $\eta(X_i)$ linearly by 
$$
\eta(X_{i}) \; \approx \;  \eta(x) + \eta^\prime(x) (X_{i} - x)  \; \equiv \;  \beta_{0} + \beta_{1} (X_{i} - x),
$$
provided that $\eta(x)$ is twice continuously differentiable.  
Estimates of $\boldsymbol{\beta} = (\beta_{0}, \beta_{1})$, and of $\eta(x)=\beta_{0}$, are then obtained by maximizing a kernel-weighted local likelihood function
\beqn \label{eq:lle}
\mathcal{L} (\boldsymbol{\beta},x) 
=  \sum_{i=1}^{n}    \ell\{ \beta_{0} + \beta_{1} (X_{i}-x) , U_{1i}, U_{2i}  \}  \:  K_{h} (X_{i}-x),
\eeqn 
where $h > 0$ is a bandwidth parameter controlling the size of the neighbourhood around $x$, $K$ is a symmetric kernel density function and $K_h (\cdot) = K(\cdot /h) /h$  weighs the contribution of each data point based on  their proximity to $x$. 
Similarly, if one uses a $p$th order local polynomial estimator, the local linear approximation in (\ref{eq:lle}) will be replaced by $\sum_{\ell=0}^p \beta_\ell (X_i-x)^\ell$ and the resulting estimator is given by $\hat{\eta}_h(x)=\hat{\beta}_0$.
Then we evaluate the log-likelihood function under the alternative hypothesis of \eqref{eq:2} as
$$
\mathbb{L}_n (\rm{H}_1,\hat \eta_{h} ) =  \sum_{i=1}^n  \ell\{ \hat{\eta}_{h} (X_i),  U_{1i}, U_{2i} \}.
$$

The difference between the two log-likelihoods allows us to evaluate the evidence in the data in favor of (or against) the null model. 
Hence, the generalized likelihood ratio statistic is given by
\beqn \label{eq:4} 
\lambda_n (h) =  \;  \mathbb{L}_n (\rm{H}_1,\hat \eta_{h} ) - \mathbb{L}_n (\rm{H}_0, \tilde \eta ).
\eeqn
While large values of $\lambda_n (h)$ suggest the rejection of the null hypothesis, we need to  determine  the rejection region for the test. In order to inform the decision in finite samples we  investigate the asymptotic distribution of the GLRT statistic under the null hypothesis.

\subsection{Asymptotic distributions of proposed GLRT statistic}

To facilitate our presentation we introduce the following notation. 
Let $f(x)>0$ be the density function of $X$ with support $\mathcal{X}$ and denote by $|\mathcal{X}|$ the range of the covariate $X$.
Also, denote by $K \ast K$ the convolution of the kernel $K$ and define
\beqn
\nonumber
\mu_n   & = &    \frac{|\mathcal{X}|}{h} \left(K(0) - \frac{1}{2} \int K^2(t) {\rm d}t \right)= \frac{|\mathcal{X}|}{h} c_K ,        \\ 
\nonumber
\nu_n    & = &    \frac{2|\mathcal{X}|}{h} \int (K(t) - \frac{1}{2} K \ast K(t))^2 {\rm d}t, \\
\nonumber
c_K&=& K(0) - \frac{1}{2} \int K^2(t) {\rm d}t.
\eeqn

The following result states that the GLRT statistic follows asymptotically a normal or equivalently a chi-square distribution in the case of negligible bias, where the mean and variance are related to the quantities $\mu_n$ and $\nu_n$, respectively.  
The technical conditions and proofs are deferred to the Appendix.

\begin{thm}
\label{glrt_thm}
Assume that the conditions {\rm (C1)}--{\rm (C7)} in the Appendix hold and the GLRT statistic $\lambda_n(h)$ is constructed from  (\ref{eq:4}) with a local linear estimator.
Then,  as $h \to 0 $ and $nh^{3/2} \to \infty$,
\beqn \label{eq:normal}
\nu_n^{-1/2} (\lambda_n(h) - \mu_n + d_n) \dlaw  N(0,1),
\eeqn
where $d_n       =     O_p(nh^4 + n^{1/2} \; h^2  )$.

Furthermore, if $\eta$ is linear or $n h^{9/2}\rightarrow 0$, then, as $n h^{3/2}\rightarrow \infty$,
\beqn \label{eq:chi2}
r_K \lambda_n(h) \asym \chi_{r_K \; \mu_n}^2,  
\eeqn
where \: $r_K = 2 \; \mu_n / \nu_n$.
\end{thm}

It should be noted that when $\eta$ is linear, the asymptotic bias $d_n$ becomes exactly zero, shown in \eqref{eq:A1} in the appendix,
and thus the condition $nh^{9/2}\rightarrow 0$ is not nedeed \citep[the optimal bandwidth for estimation is of the order $n^{-1/5}$, see][]{Acar/Craiu/Yao:2011}.
More importantly, this facilitates the calculation of the GLRT statistic $\lambda_n(h)$ in practice, since one can use directly  the bandwidth used for estimation, chosen by the leave-one-out cross-validated likelihood \citep{Acar/Craiu/Yao:2011}. Our simulation study in Section 3 provides empirical support for this suggestion. 
 
Moreover, the asymptotic results in Theorem \ref{glrt_thm}  can be easily extended to the case where  $\lambda_n(h)$ is based on a $p$th order local polynomial estimator, by substituting the kernel function $K$ with its equivalent kernel $K^\ast$ in $c_K$ and $r_K$ \citep[see][page 64, for the expression of $K^\ast$]{Fan/Gijbels:1996}  induced by the local polynomial fitting \citep{Fan/Zhang/Zhang:2001}.  
The asymptotic chi-square distribution (\ref{eq:chi2}) continues to  hold if either $\eta$ is a polynomial
of degree $p$ or $n h^{(4p+5)/2}\rightarrow 0$, as the asymptotic bias $d_n=O_p(nh^{2p+2}+n^{1/2} h^{p+1})$. The practical implication of such an extension is that, if the interest is to test a null hypothesis of a polynomial form $\eta(x)=\sum_{\ell=0}^p \beta_\ell x^\ell$, it is recommended to calculate $\lambda_n(h)$ using the local polynomial estimator with the corresponding degree $p$. This avoids the possible necessity of undersmoothing in order to have the asymptotic bias negligible. 

As pointed out earlier, the hypothesis of $\eta$ being constant is a special case of the linearity constraint and leads to the classical copula model
(i.e., no covariate adjustment is required). If this hypothesis is of interest, using a local constant estimator, i.e., $p=0$, to calculate $\lambda_n(h)$ may be more appealing  (as confirmed by the simulations in Section 3) than using a local linear estimator.
The latter tends to overfit even with large bandwidth when $H_0$ indeed holds, thus resulting in an inflated type I error.

One can conclude from Theorem \ref{glrt_thm} that the GLRT is fairly similar to the classical likelihood ratio test. 
The tabulated value of the scaling constant $r_K$ is close to $2$ for commonly used kernels. For instance, $r_K = 2.115 $ for the commonly used Epanechnikov kernel $K(u)=0.75 (1-u^2) \mathbf{1}_{\{|u|\le 1\}}$.
The degrees of freedom (df) $r_K \; c_K \;  |\mathcal{X}| /h$ of the asymptotic null distribution of the GLRT tends to infinity when $h\rightarrow 0$, due to the nonparametric nature of the alternative hypothesis. One can interpret the quantity $ |\mathcal{X}| /h$ as the number of nonintersecting intervals on $\mathcal{X}$, and thus $r_K \; c_K \;  |\mathcal{X}| /h$ approximates the effective number of parameters in the nonparametric estimation.
For the Epanechnikov kernel with $c_K = 0.45$, the degrees of freedom is given by $0.968\; |\mathcal{X}| /h$.

\section{Simulation Study}
\label{s:sec3}

We conduct simulations to 
evaluate the finite sample performance of the proposed test for the linear hypothesis given in (\ref{eq:2}).
We consider three simulation scenarios corresponding to three calibration functions,  
\bea
\mbox{$\mathbf{M}_0$}: \quad \eta_0(X)&=& 8, \\
\mbox{$\mathbf{M}_1$}: \quad \eta_1(X)&=& 25- 4.2 \: X, \\
\mbox{$\mathbf{M}_2$}: \quad \eta_2(X) &=&  12 + 8\sin(0.4\:X^2).
\eea
The copula used  belongs to   the Frank family and has the form
$$
C(u_{1},u_{2}|\theta) =  - \frac{1}{\theta}  \ln \left\{  1+ \frac{(e^{-\theta u_{1}}-1)    (e^{-\theta u_{2}}-1)}{   e^{-\theta }-1} \right\} , \quad \theta \in (-\infty,\infty) \setminus \{0\}.
$$
Since the range of $\theta$ is $\RR \backslash \{0\}$, an identity link is used, i.e., $\theta_k(X)= \eta_k(X)$ for $k=0,1, 2$. Similar findings (not reported here) were obtained for the simulations using the Clayton copula.

Our Monte Carlo experiment consists of $200$ replicated samples of sizes  $n=200$ and $500$ generated from each model.  Specifically, under  model $\mathbf{M}_k$ we first simulate the covariate values $X_i \sim \mbox{U}(2,5)$, $i=1,\ldots, n$  and then, conditional on $X_i$, the uniform pairs  $(U_{1i}, U_{2i})$  are sampled from the Frank family with  copula parameter $\theta_k(X_i)=\eta_k(X_i)$ induced by the calibration model $\mathbf{M}_{k}$, for all $k=0,1, 2$. 
Throughout the simulations we have used the Epanechnikov kernel.  
For each Monte Carlo sample, the leave-one-out cross-validated likelihood method of \cite{Acar/Craiu/Yao:2011} is employed to select, out of 12 pilot values ranging from
0.33 to 2.96 and equally spaced in logarithmic scale, the optimum  bandwidth $h$ for the local polynomial estimation of the calibration function. 
We have  followed the suggestion made in Section  2 and have calculated the nonparametric estimator for $\eta$ using a local polynomial of the same degree as specified by the null hypothesis. 
For instance, in Table 1, when testing $H_0: \eta=c$, we consider a local constant estimator (with $p=0$) for $\eta$ under the alternative model. 
Subsequently, the GLRT statistic $\lambda_n(h)$ is computed using the same bandwidth $h$ that is used for estimation.  
We also assume that in practice one would {\it first} test for constant calibration function and, conditional on rejection, would test for  linear calibration. 
For this reason, in Table 1 we do not report the results of testing $H_0: \eta(x)=a_0+a_1x$ when the generating model is $\mathbf{M}_0$.

\begin{table*}[h!] 
\centering
\caption{Demonstration of the proposed GLRT for testing the linear/constant null hypothesis $H_0$ at $\alpha=0.10, 0.05$ and $0.01$, respectively. Shown are the rejection frequencies assessed from 200 Monte Carlo replicates. The  sample  sizes are $n=200$ and $n=500$, where the generating models are shown in the  ``True Model'' column. Those entries in the table reflecting the power of the testing procedure are shown in bold face.}
\label{table1}               
\vspace{0.1in}                                                                                                       
\begin{tabular}{ccc    ccc   cc   ccc }                                                                                   
\hline \hline \\ [-3.5ex]          
&&& \multicolumn{7}{c}{Null Model} \\       [-0.5ex]                                                                                                                                                                                                                                                                &&& \multicolumn{3}{c}{{$H_0: \eta(x)=a_0+a_1 x$}}     && \multicolumn{3}{c}{{$H_0: \eta=c$}}  \\
      \cline{4-6} \cline{8-10}                                                                                                                                            
{True Model}   & $n$ &&  .10 &  .05   &  .01   &&   .10 &  .05   &  .01 \\                                                                             
\hline                                                    
  & $200$ &&   --- &  --- &  ---         && .105 & .040 & .020  \\
\raisebox{2ex}[0pt]{$\mathbf{M}_0$} 
& $500$   && --- & --- & ---             && .110 & .045 & .005  \\
\hline
 & $200$ &&      .100 &  .055 & .005   && {\bf .995} & {\bf .990} & {\bf .955} \\
 \raisebox{2ex}[0pt]{$\mathbf{M}_1$} & 
 $500$   &&  .085 & .055 & .010       && {\bf 1.00} & {\bf 1.00} & {\bf 1.00} \\                                                            
 \hline                                                   
 & $200$ &&   {\bf 1.00} &  {\bf 1.00} &  {\bf 1.00}  && {\bf 1.00} & {\bf 1.00} & {\bf 1.00} \\
 \raisebox{2ex}[0pt]{$\mathbf{M}_2$} 
 & $500$  && {\bf 1.00} &  {\bf 1.00} &  {\bf 1.00}  && {\bf 1.00} & {\bf 1.00} & {\bf 1.00}\\      
 \hline
\hline                                                                                                                                                         
\end{tabular}    
\end{table*}

One can notice from Table \ref{table1} that the rejection rates under the null are very close to the target values of the type I error probabilities $\alpha \in \{0.1, 0.05,0.01\}$, for both linear and constant nulls (models $\mathbf{M}_0$ and $\mathbf{M}_1$). 
Our approach leads to high power in detecting departures from the null, as one can see from the results  generated under models  $\mathbf{M}_1$ and $\mathbf{M}_2$. For clearer visualization, the entries in the table that correspond to power are shown in bold face.

\section{Data Application}
\label{s:sec4}

In this section, we apply the GLRT to the two data examples studied in \cite{Acar/Craiu/Yao:2011}. 
Our aim is to check whether a constant copula model or a conditional copula model with a linear calibration function fits these examples reasonably well, i.e. whether the nonparametric calibration estimates of  \cite{Acar/Craiu/Yao:2011} are in fact necessary.

\subsection{Twin birth data}

\begin{figure}[h!]
\centering
\includegraphics[width=\textwidth]{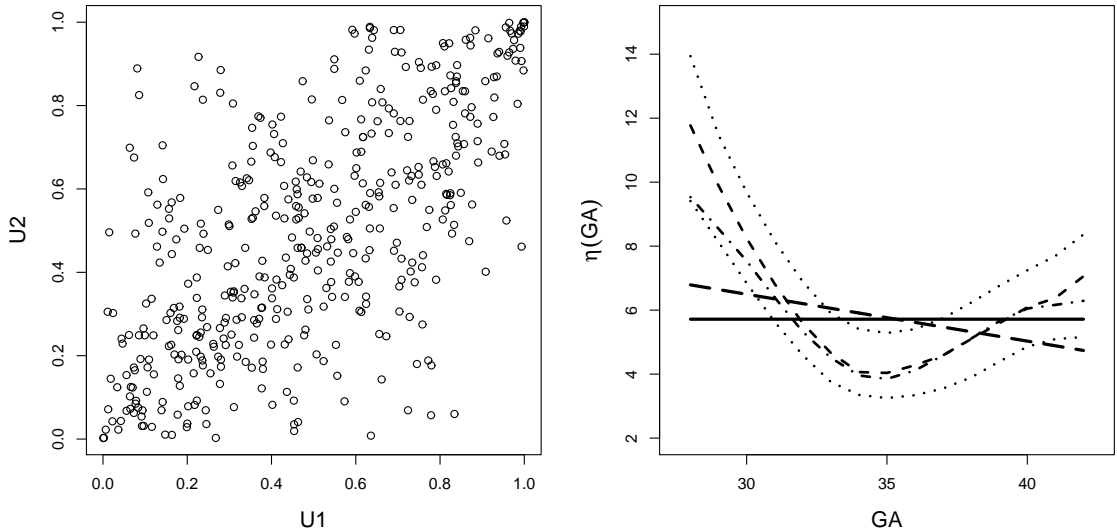}
\caption{Scatterplot of the conditional marginal distributions of birth weights given the gestational age (left panel) and the plot of calibration function estimates under the Frank copula (right panel): maximum likelihood estimate of the constant calibration function (solid line), maximum likelihood estimate of the linear calibration function (long-dashed line), local constant estimates (dot-dashed line), local linear estimates (dashed line), $90\%$ pointwise confidence intervals for the local linear estimates (dotted lines).\label{fig:1}}
\end{figure}

This data set contains the birth weights and the gestational age of 450 twin pairs from the Matched Multiple Birth Data Set (MMB) of the National Center for Health Statistics. 
Of interest is the dependence between the birth weights (BW$_1$, BW$_2$) of the first- and second-born twins given their gestational age GA.
We follow \cite{Acar/Craiu/Yao:2011} and transform the data on the uniform scale, as shown in the left panel of Figure \ref{fig:1}, and use the Frank family of copulas to model the dependence structure.  The right panel of Figure \ref{fig:1} shows the maximum likelihood estimates obtained under the constant calibration assumption (solid line), linear calibration assumption (long-dash line), the nonparametric estimates with $p=0$ (dot-dashed line), $p=1$ (dashed line) and $90\%$ pointwise confidence intervals for the local linear estimates (dotted lines), obtained as in \cite{Acar/Craiu/Yao:2011}.

As seen in Figure \ref{fig:1}, the maximum likelihood estimates under constant and linear calibration assumptions are not within the confidence intervals of the local linear estimates, suggesting that these simple parametric formulations may not be appropriate. This empirical observation is confirmed by the GLRT tests, which yielded p-values smaller than $10^{-5}$ for both tests (test statistics are 13.58 on $3.92$ df and 12.95 on $3.36$ df for the constant and linear hypothesis, respectively).

Thus, we conclude that the variation in the strength of dependence between the twin birth weights at different gestational ages, as represented by the nonparametric estimates in the right panel of Figure \ref{fig:1} is statistically significant.

\bigskip

\subsection{Framingham Heart Study data}

\begin{figure}[h!]
\centering
\includegraphics[width=\textwidth]{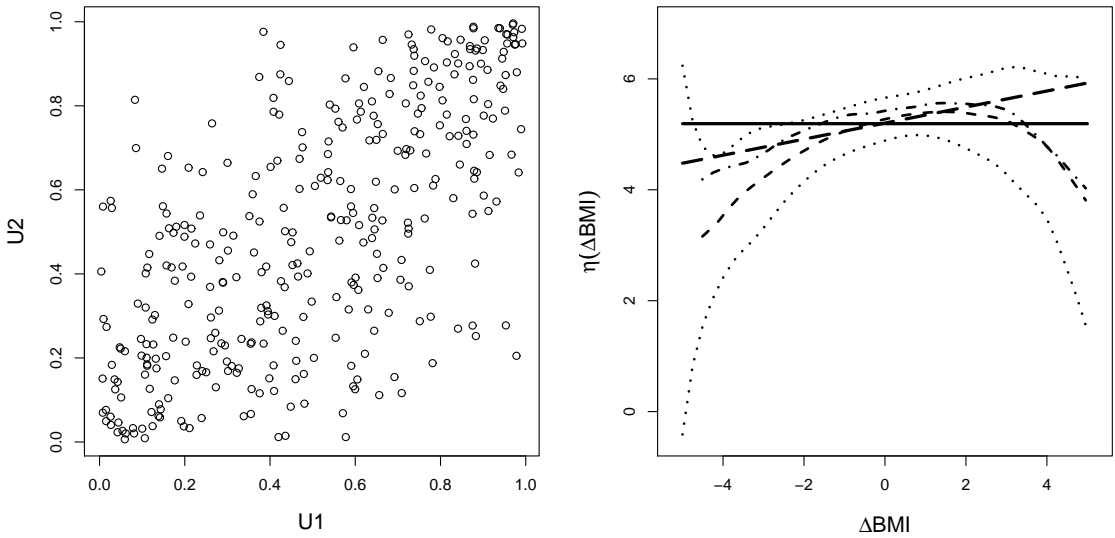}
\caption{Scatterplot of the conditional marginal distributions of the log-pulse pressures given the change in body mass index (left panel) and the plot of calibration function estimates under the Frank copula (right panel): maximum likelihood estimate of the constant calibration function (solid line), maximum likelihood estimate of the linear calibration function (long-dashed line), local constant estimates (dot-dashed line), local linear estimates (dashed line), $90\%$ pointwise confidence intervals for the local linear estimates (dotted lines). \label{fig:2}}
\end{figure}

This data set comes from the Framingham Heart Study (FHS) and contains the log-pulse pressures of 348 subjects at the first two examination periods, denoted by $\log(PP_1)$ and $\log(PP_2)$, respectively, as well as the change in body mass index $\Delta \text{BMI}$ between these periods. 
The left panel of Figure \ref{fig:2} displays the conditional marginal distributions of the log-pulse pressures given $\Delta \text{BMI}$, which are obtained parametrically as in \cite{Acar/Craiu/Yao:2011}. 

The estimates of the calibration function are obtained under the chosen Frank family using the maximum likelihood estimation with constant and linear calibration forms and the nonparametric estimation with $p=0$ and $p=1$. The results are shown in the right panel of Figure \ref{fig:2}. 

Based on the Figure \ref{fig:2} we suspect that a constant copula model may be appropriate. 
To decide whether the fitted constant copula model is appropriate, we perform the GLRT using the local constant estimates at the bandwidth value $h=3.45$. This bandwidth choice leads to 2.66 df of the chi-square distribution.
The difference between the log-likelihoods of the alternative and null conditional copula models is 0.91 and consequently the p-value is 0.514. Thus, we conclude that the change in body mass index does not have any significant effect on the strength of dependence between the two log-pulse pressures.

\section{Conclusion}
\label{s:sec:5}

Adjusting statistical dependence for covariates  via conditional copulas is an active area of research where model fitting and validation are currently in early development.
This paper takes a first step towards establishing conditional copula model diagnostics by presenting a formal test of hypothesis for the calibration function.
Inspired by the generalized likelihood ratio idea of \cite{Fan/Zhang/Zhang:2001}, the proposed test uses the local likelihood estimator of \cite{Acar/Craiu/Yao:2011} to specify the model under the alternative when testing a parametric calibration function hypothesis. 
The asymptotic null distribution of the test statistic, shown to be a chi-squared distribution with the number of degrees of freedom determined by  the estimation-optimal bandwidth, is used to determine the rejection region in finite samples. Simulations suggest that the method has high power of
detecting departures from the null model and  yields the targeted type I error probability.

The GLRT procedure presented here can be easily adapted to test an arbitrary parametric calibration function.
Furthermore, the approach can be extended to employ other nonparametric estimators, such as smoothing splines, although with additional effort of deriving the asymptotic null distribution. 
Nevertheless, the asymptotic null distribution may not  always be appropriate for determining the rejection region in finite samples. 
While conditional bootstrap is usually used to assess the null distribution of the GLRT in regression-based problems, defining a similar bootstrap procedure in the conditional copula setting is not straightforward and requires further study.

\bibliography{GLRT}
\bibliographystyle{asa}

\renewcommand{\thesection}{\Alph{section}}
\setcounter{section}{0} 

\section{Regularity Conditions and Technical Proofs} 

The asymptotic distribution of the GLRT statistic relies on the following technical conditions. The conditions (C1)-(C3) are standard in nonparametric estimation and the conditions (C4)-(C7) are required to regularize the conditional copula density.

\begin{list}{\labelitemi}{\leftmargin=2em}
\item[(C1)] The density function $f(X) > 0$ of the covariate $X$ is Lipschitz continuous, and X has a bounded support $\mathcal{X}$.
\item[(C2)] The kernel function $K(t)$ is a symmetric probability density function that is bounded and Lipschitz continuous. 
\item[(C3)] The  functions $\eta$ and $g^{-1}$ have $(p+1)$th continuous derivatives, where $p=1$ when a local linear estimator is used for $\lambda_n(h)$.

\item[(C4)] The functions $\ell_{1} \{  \eta(x),u_{1},u_{2} \} $ and $ \ell_{2}\{  \eta(x), u_{1},u_{2} \}$ exist and are continuous on $\mathcal{X} \times (0,1)^{2}$, and can be bounded by integrable functions of $u_{1}$ and $u_{2}$.
\item[(C5)] $\rm{E} \big| \{ \ell_1 (\eta(x), u_1,u_2) \mid x \}  \big|^4 < \infty $.
\item[(C6)] $\rm{E} \{  \ell_2 (\eta(x), u_1,u_2)\mid x \} $ is Lipschitz continuous.
\item[(C7)] The function $\ell_2 (t, u_1,u_2) <0 $ for all $t \in \mathbb{R}$, and $u_1,u_2 \in (0,1)$. For some integrable function $k$, and for $t_1$ and $t_2$ in a compact set,
$$ | \ell_2 (t_1, u_1,u_2) - \ell_2 (t_2, u_1,u_2)|  < k(u_1,u_2) |t_1-t_2|.$$ 
In addition, for some constants $\xi >2$ and $k_0>0$, $j=1, 2, 3$,
\bea 
E\Big\{ \mysup_{x, || \bs{m} || < k_0/\sqrt{nh}}    | \ell_2 (\bar \eta(x,X) + \bs{m}^T  \bs{z}_{x},  U_{1}, U_{2}  )|   \Big|   \frac{X-x}{h} \Big|^{j-1} 
  K \Big(\frac{X-x}{h}\Big)     \Big\} ^\xi 
= O(1),
\eea
 where $\bar \eta(x,X) = \eta(x) + \eta' (x) (X-x)$.
\end{list}
\bigskip

Before proving Theorem \ref{glrt_thm}, we shall introduce additional notation.
Let $\gamma_n = 1/ \sqrt{nh}$
and
define
\beqn 
\nonumber
\alpha_n(x)&    =  &  \frac{\gamma_n^2 }{\sigma^2(x) f(x)} \sum_{i=1}^n \ell_1(\eta(X_i),U_{1i},U_{2i}) \;  K((X_i - x)/h),    \\ \nonumber
R_n(x) &=&      \frac{ \gamma_n^2 }{\sigma^2(x) f(x)}    \sum_{i=1}^n  \Big\{ \ell_1(\bar \eta(x, X_i),U_{1i},U_{2i})  - \ell_1(\eta(X_i),U_{1i},U_{2i}) \Big\}  \times K((X_i - x)/h),
\eeqn
where 
$\sigma^2(x)= -\rm{E} \big[  \; \ell_2 \; \{ \eta(x) , U_{1},U_{2} \} \; | \;X=x    \big]  $ denotes the Fisher Information for $ \eta(x)$ at any $x \in \mathcal{X}$.

Recall that $\bar \eta(x,X_i) = \eta(x) + \eta' (x) (X_i-x)$, define
\beqn
\nonumber      R_{n1}  &=&  \sum_{k=1}^n \ell_1(\eta(X_k),U_{1k},U_{2k})    \; R_n(X_k),  \\
\nonumber      R_{n2}  &=&  -\sum_{k=1}^n \ell_2(\eta(X_k),U_{1k},U_{2k}) \;  \alpha_n(X_k)\; R_n(X_k), \\
\nonumber      R_{n3}  &=&  -\frac{1}{2} \sum_{k=1}^n  \ell_2(\eta(X_k),U_{1k},U_{2k}) \;  R_n^2(X_k). 
\eeqn
and set
\beqn \nonumber
T_{n1}    &=&         \gamma_n^2   \; \sum_{i=1}^n  \sum_{k=1}^n      \frac{    \ell_1(\eta(X_k),U_{1k},U_{2k})     }{  \sigma^2(X_k) f(X_k)}  \;  \ell_1(\eta(X_i),U_{1i},U_{2i})      \;  K((X_i - x)/h),  
  \\ \nonumber 
  T_{n2}   & = &       \gamma_n^4  \;     \sum_{i=1}^n   \sum_{j=1}^n   \;   \ell_1(\eta(X_i),U_{1i},U_{2i}) \;  \ell_1(\eta(X_j),U_{1j},U_{2j})   \\ \nonumber 
&& \qquad \qquad 
\times \left\{  \; \sum_{k=1}^n  \frac{    \ell_2(\eta(X_k),U_{1k},U_{2k})    }{ ( \sigma^2(X_k) f(X_k) )^2}     \;  K((X_i - x)/h)   K((X_i - x)/h) \right\}. 
\eeqn

\bigskip
\noindent The Lemma \ref{lemma:1}--\ref{lemma:3} are used in our derivations, and their proofs are given at the end of this appendix.

\begin{lem}
\label{lemma:1}
Under conditions {\rm (C1)}--{\rm (C7)},
$$  \hat{\eta}_h (x)    -   \eta(x) = \{  \alpha_n (x)   + R_n (x)\} \; (1+o_p(1)). $$
\end{lem}

\begin{rem}
Note that, when $\eta$ is linear, then $R_n (x)$ directly becomes zero as for each $i= 1,\ldots,n$
\begin{AEquation}\label{eq:A1}
\bar \eta(x, X_i) = a_0+a_1x + a_1(X_i-x) = \eta(X_i). 
\end{AEquation}
This is clearly also the case when $\eta$ is constant.
\end{rem}

\bigskip
\begin{lem}
\label{lemma:2}
Under conditions {\rm (C1)}--{\rm (C7)}, as $h \to 0$ and $nh^{3/2} \to \infty$
\begin{multline*}
T_{n1}   =  \frac{1}{h} K(0)  \rm{E} [f^{-1}(X) ]       + \frac{1}{n} \sum_{k\neq i}
   \frac{ \ell_1(\eta(X_k),U_{1k},U_{2k}) }{  \sigma^2(X_k) f(X_k)}   \; \ell_1(\eta(X_i),U_{1i},U_{2i})  \\
  \times K_h\left(X_i - X_k\right) + \;  o_p(h^{-1/2}) ,  
\end{multline*}
\vspace{-0.2in}
\begin{multline*}
T_{n2}   =   - \;  \frac{1}{h} \rm{E} [f^{-1}(X)] \int K^2(t) dt 
  \;  - \;   \frac{2}{nh}    \sum_{i < j}    \frac{ \ell_1(\eta(X_i),U_{1i},U_{2i}) }  {   \sigma^2(X_i) f(X_i)  }   \\
 \times  \ell_1(\eta(X_j),U_{1j},U_{2j})    K \ast K(( X_j- X_i) /  h ) + o_p(h^{-1/2}).
\end{multline*}
\end{lem}
\bigskip

To introduce Lemma \ref{lemma:3}, we first restate a proposition in  \cite{Jong:1987}, where the notation is adapted to ours.
Let $X_1,X_2, \ldots$ be independent variables, and $w_{ijn}(\cdot,\cdot)$ Borel functions such that 
$W(n)= \sum_{1\leq i \leq n}   \sum_{1\leq j \leq n}  w_{ijn}(X_i,X_j),$
and $W_{ij}= w_{ijn}(X_i,X_j)$ $ +\; w_{jin}(X_j,X_i)$, where
the index $n$ is suppressed in $W_{ij}$.
Following de Jong (1987, Definition 2.1), 
 $W_n$ is called clean if the conditional expectations of $W_{ij}$ vanish:
$E[W_{ij}| X_i] = 0 \  a.s. \ \text{for all} \quad i,j \leq n.$

\bigskip
\noindent {\bf Proposition 3.2} \citep{Jong:1987} \ \ {\em
Let $W(n)$ be clean with variance $\nu_n^\ast$, if $G_{I}$, $G_{II}$ and $G_{IV}$ be of lower order than $\nu_n^{\ast 2}$, then 
\beq \nonumber
\nu_n^{\ast -1/2} W(n) \dlaw N(0,1), \qquad n \to \infty,
\eeq
where
\beqn \nonumber G_{I} &=&  \sum_{1\leq i <j \leq n} \E (W_{ij}^4),  \hspace{0.15in} G_{II} =  \sum_{1\leq i <j <k \leq n} \{ \E (W_{ij}^2 W_{ik}^2   )+ \E (W_{ji}^2 W_{jk}^2   ) +\E (W_{ki}^2 W_{kj}^2   ) \}, \\ \nonumber G_{IV} &=&  \sum_{1\leq i <j <k <l  \leq n} \{ \E (W_{ij} W_{ik} W_{lj} W_{lk}    )   +\E (W_{ij} W_{il} W_{kj} W_{kl}    ) + \E (W_{ik} W_{il} W_{jk} W_{jl}    )         \}.\eeqn
 }

We now define the following U-statistic,
\begin{AAlign}\label{eq:A2}
W(n) &= \frac{\sqrt{h}}{n} \sum_{i \neq j}      \frac{1}{(\sigma^2(X_i) f(X_i))^2}    \ell_1(\eta(X_j),U_{1j},U_{2j})\; \ell_1(\eta(X_i),U_{1i},U_{2i}) \; \nonumber \\
& \times \{ 2 K_h(X_j- X_i) - K_h \ast K_h(X_j- X_i ) \}.
\end{AAlign}

\bigskip

\begin{lem}
\label{lemma:3}
Under conditions {\rm (C1)}--{\rm (C7)}, $W_n$ defined in (\ref{eq:A2}) is clean
and $W(n) \dlaw N(0, \nu^\ast)$,  as $h\rightarrow 0$ and $nh^{3/2}\rightarrow \infty$,
where $\nu^\ast = 2 \;|| 2K - K\ast K||_2^2  \;E[ f^{-1}(X)]$. 
\end{lem}

\bigskip

\noindent {\bf Proof of Theorem \ref{glrt_thm}}.\ \ 
To provide a general framework, we use $\eta(X_k)$ and $\tilde \eta(X_k)$ to denote the true value 
under the null hypothesis and its maximum likelihood estimator, respectively. 
Then, the GLRT statistic can be written as
\beqn
\nonumber 
 \lambda_n (h) &=&    \sum_{k=1}^n  [  \ell (  \hat{\eta}_h (X_k) ,   U_{1k}, U_{2k}  ) - \ell (   \eta(X_k) ,   U_{1k}, U_{2k}  ) \\
\nonumber 
&&\qquad \qquad\qquad  - \{ \ell (   \tilde \eta(X_k),   U_{1k}, U_{2k}  ) - \ell (   \eta(X_k) ,   U_{1k}, U_{2k}  ) \}]
 \\
\nonumber
& \equiv &  \lambda_{1n} (h) -  \lambda_{2n}.
\eeqn
Here $\lambda_{2n}$ corresponds to the canonical likelihood ratio statistic
and it is $\lambda_{1n} (h)$ that governs the asymptotic distribution of $ \lambda_n (h)$.

To derive the asymptotic distribution of $\lambda_{1n} (h)$, first approximate 
$\ell (   \hat{\eta}_h (X_k)   ,   U_{1k}, U_{2k}  )$ around $\eta(X_k)$
\beqn    
\nonumber
\lambda_{1n} (h) &\approx&       \sum_{k=1}^n 
\ell_1 (    \eta(X_k),   U_{1k}, U_{2k}  )   \;  \{   \hat{\eta}_h (X_k)    -  \eta(X_k) \} \\ \nonumber
 & & \qquad   +     \frac{1}{2} \;  \sum_{k=1}^n \;     \ell_2 (   \eta(X_k),   U_{1k}, U_{2k}  )   \;  \{    \hat{\eta}_h (X_k)    -   \eta(X_k) \} ^2.
\eeqn
Applying Lemma \ref{lemma:1} and Lemma \ref{lemma:2} yields
\beqn
\nonumber
- \lambda_{1n}(h) &=&  -  \: h^{-1}  \rm{E} [f^{-1}(X)]  \left\{  K(0)  -   \int K^2(t) dt /2\right\} \\ \nonumber
&& -  \:n^{-1} \sum_{i\neq j}
   \frac{ \ell_1(\eta(X_i),U_{1i},U_{2i})}{(\sigma^2(X_i) f(X_i))^2}    \;    \ell_1(\eta(X_j),U_{1j},U_{2j})  K_h\left(X_j - X_i\right) \\ \nonumber
&& + \: n^{-1}    \sum_{i < k}      \frac{ \ell_1(\eta(X_i),U_{1i},U_{2i})}{(\sigma^2(X_i) f(X_i))^2}    \;   \ell_1(\eta(X_j),U_{1j},U_{2j})  K_h \ast K_h ( X_j- X_i) \\ \nonumber
&&  - \: R_{n1} +  R_{n2} +  R_{n3} +O_p\left(n^{-1}h^{-2}\right) + o_p(h^{-1/2}).
\eeqn
By calculating of the leading terms $R_{n1} $, $R_{n2}$ and $R_{n3}$, one can show that
\beqn
\nonumber
R_{n1} &=& 
\sum_{k=1}^n  \frac{h^2}{2} \; \ell_1 (\eta(X_k),U_{1k},U_{2k})  \eta''(X_k) \int t^2 \; K(t) dt  (1+ o_p(1))
=O_p(n^{1/2}h^2), 
\\ \nonumber
- R_{n2} &=&  \sum_{k=1}^n  \frac{h^2}{4} \;    \frac{\ell_1 (\eta(X_k),U_{1k},U_{2k}) }{\sigma^2(X_k) f(X_k)}   \eta''(X_k) \; \omega_0 (1+ o_p(1))   =  O_p(n^{1/2}h^2),
\\ \nonumber
- R_{n3} &=&  \frac{n h^4}{8} E \eta''(X)^2  \sigma^2(X)\;  \omega_0 (1+o_p(1)) = O_p(n h^4),
\eeqn
where $ \omega_0= \int \int t^2 (s+t)^2 K(t) K(s+t) \;ds \;dt$. 
Thus, 
$$R_{n3} - (R_{n1}-R_{n2})= O_p(nh^4 + n^{1/2}h^2).$$ 
This results in
\beq
\nonumber
- \lambda_{1n}(h) = - \mu_n + d_n -h^{-1/2}\; W(n)/2  + o_p(h^{-1/2}),
\eeq
where $W_n$ is as defined in (\ref{eq:A2}). Applying Lemma \ref{lemma:3}, we arrive at 
$W(n) \dlaw N(0, \nu^\ast),$  
where $\nu^\ast = 2 \;|| 2K - K\ast K||_2^2  \;E[ f^{-1}(X)]$. 
Hence,
$$  \nu_n^{-1/2} ( \lambda_{1n}(h)  - \mu_n + d_n )   \dlaw  N(0, 1),$$
where $ \nu_n =(4h)^{-1} \nu^\ast  $. 
For the asymptotic null distribution of $\lambda_{n}(h)$, this result can be re-written as
$$  \nu_n^{-1/2} \{ ( \lambda_{1n}(h) - \lambda_{2n}) - \mu_n + d_n + \lambda_{2n} \}   \dlaw  N(0, 1). $$
Since $\lambda_{2n} = O_p(1)$, it vanishes compared to $\lambda_{1n}(h) = O_p(h^{-1})$ and we obtain
$$  \nu_n^{-1/2} ( \lambda_{n}(h) - \mu_n + d_n )   \dlaw  N(0, 1). $$

For the second result, note that the distribution $N(a_n, 2a_n)$ is approximately same as the chi-square distribution with degrees of freedom $a_n$, for a sequence $a_n \to \infty$. Letting $a_n= 2 \mu_n^2 / \nu_n$ and $r_K= 2 \mu_n / \nu_n$, we have
$$ (2 a_n)^{-1/2}   ( r_K \lambda_n(h)  - a_n )   \dlaw  N(0, 1),$$
provided that $d_n$ vanishes.
\hfill $\Box$

\subsection*{Additional Technical Details}

\noindent {\bf Proof of Lemma \ref{lemma:1}}.
Define
$$ 
\bs{b} = \gamma_n^{-1} ( \beta_0 - \eta(x), h (\beta_1- \eta'(x)))^T, 
$$
so that each component has the same rate of convergence. 
Then, we have
$$ 
\beta_0 + \beta_1 (X_i -x)= \bar  \eta(x,X_i) + \gamma_n \bs{b}^T \bs{z}_{i,x}, 
$$
where $ \bs{z}_{i,x} = (1, (X_i -x)/h)^T$.
The local log-likelihood function can be re-written in terms of $\bs{b}$,
$$  
\mathcal{L}(\bs{b}) = \sum_{i=1}^n \ell (\bar \eta(x,X_i) +\gamma_n \bs{b}^T \bs{z}_{i,x}, U_{1i}, U_{2i}  ) K_h(X_i-x).
$$
Note that $\hat{\bs{b}}=  \gamma_n^{-1} ( \hat{\beta}_0 - \eta(x), h (\hat{\beta}_1- \eta'(x)))^T$ maximizes $ \mathcal{L}(\bs{b})$. 
It also maximizes following normalized function,
$$
 \mathcal{L^\ast}(\bs{b}) =  \sum_{i=1}^n \Big\{  \ell (\bar \eta(x,X_i) +\gamma_n \bs{b}^T \bs{z}_{i,x}, U_{1i}, U_{2i}  ) -\ell (\bar \eta(x,X_i) , U_{1i}, U_{2i}  ) \Big\} K((X_i - x)/h),
$$
which can be written as
\beqn
\nonumber
\mathcal{L^\ast}(\bs{b}) 
 &=& h \gamma_n \sum_{i=1}^n \ell_1(\bar \eta(x,X_i) , U_{1i}, U_{2i}  )\; \bs{b}^T \bs{z}_{i,x} K_h(X_i -x)  \\ \nonumber
 && +\;  h \frac{ \gamma_n^2}{2} \sum_{i=1}^n \ell_2 (\bar \eta(x,X_i) + \bs{m_n}^T  \bs{z}_{i,x},  U_{1i}, U_{2i}  )\; ( \bs{b}^T \bs{z}_{i,x} )^2 \;  K_h(X_i -x) \\ \nonumber
&=&  \bs{b}^T \Big\{  \gamma_n \sum_{i=1}^n \ell_1 (\bar \eta(x,X_i) , U_{1i}, U_{2i}  )  \bs{z}_{i,x} K((X_i - x)/h) \Big\} \\ \nonumber
&&   +\;  2^{-1}  \bs{b}^T    \Big\{   \frac{1}{n} \sum_{i=1}^n \ell_2 (\bar \eta(x,X_i) + \bs{m_n}^T  \bs{z}_{i,x},  U_{1i}, U_{2i}  )\;  \bs{z}_{i,x}   \bs{z}_{i,x}^T \;  K_h(X_i -x) \Big\} \bs{b}.
\eeqn
In the following, we will show that
\beq
\nonumber
n^{-1} \sum_{i=1}^n \ell_2 (\bar \eta(x,X_i) + \bs{m_n}^T  \bs{z}_{i,x},  U_{1i}, U_{2i}  )\; \bs{z}_{i,x}  \bs{z}_{i,x}^T \;  K_h(X_i -x)  =  - \Delta + o_p(1), 
\eeq
where
$ \Delta =    \sigma^2(x) f_X(x) \left( \begin{array}{cc} \mu_0, & \mu_1   \\ \mu_1,  & \mu_2 \end{array} \right), $
with $\mu_i=\int t^i K(t) dt$, and $o_p(1)$ is uniform in $x \in \mathcal{X}$ and $||\bs{b}|| < m_0 $, for some fixed constant $m_0 > 0$. 
To show this, we need the following smoothness result.
Let $ A_n (x,\bs{m}) = \ell_2 (\bar \eta(x,X) + \bs{m}^T  \bs{z}_{x},  U_{1}, U_{2}  )\; \bs{z}_{x}     \bs{z}_{x}^T    \;  K_h(X-x),$
with $|| \bs{m}|| < 1$. 
Then, under the conditions (C1)-(C7), we can show that
$$
 | A_n (x_1,\bs{m_1}) - A_n (x_2,\bs{m_2})| \;   \leq h^{-3}  \; k(X,U_1,U_2)  (|| \bs{m_1}-\bs{m_2}||  + |x_1-x_2|)
$$
for some integrable function $k(X,U_1,U_2)$. 
Thus, using the triangle inequality,
\bea
&&\hspace{-0.28in} \Bigg| \frac{1}{n} \sum_{i=1}^n \ell_2 (\bar \eta(x,X_i) + \bs{m_n}^T  \bs{z}_{i,x},  U_{1i}, U_{2i}  )\; \bs{z}_{i,x}  \bs{z}_{i,x}^T \;  K_h(X_i -x) - (- \Delta)  \Bigg|  
\\ 
&&   
\hspace{-0.28in} \leq   \;  \frac{1}{n} \sum_{i=1}^n   \Big| \{ \ell_2 (\bar \eta(x,X_i) + \bs{m_n}^T  \bs{z}_{i,x},  U_{1i}, U_{2i}  )- \ell_2 (\bar \eta(x,X_i),  U_{1i}, U_{2i}  ) \}  \bs{z}_{i,x}  \bs{z}_{i,x}^T   K_h(X_i -x) \Big|
\\ 
&&
\hspace{0.09in} +  \mysup_{\eta, x}      \Bigg[       \frac{1}{n}    \Big|  \sum_{i=1}^n \{ \ell_2 (\bar \eta(x,X_i) ,  U_{1i}, U_{2i}  ) - \ell_2 (\eta(X_i) , 
 U_{1i}, U_{2i}  ) \}  \times \bs{z}_{i,x}  \bs{z}_{i,x}^T \;  K_h(X_i -x) \Big| \Bigg] 
 \\
 &&
\hspace{0.3in} +   \mysup_{\eta, x}      \Bigg[    \Big|  \frac{1}{n} \sum_{i=1}^n \ell_2 (\eta(X_i) ,  U_{1i}, U_{2i}  )  
 \bs{z}_{i,x}  \bs{z}_{i,x}^T \;  K_h(X_i -x)  - E \{  \ell_2 (\eta(X) ,  U_{1}, U_{2}  )  \bs{z}_{x}  \bs{z}_{x}^T
 \\
 && 
\hspace{1in}  \times \;  K_h(X-x) |x\}  \Big| 
+  \Big|  E \{  \ell_2 (\eta(X) ,  U_{1}, U_{2}  ) \; \bs{z}_{x}  \bs{z}_{x}^T \;  K_h(X-x) |x\}  + \Delta \Big| \Bigg],
\eea
for $\eta$ in a compact set and $x \in \mathcal{X}$. The first sum goes to zero by the previous argument and the Dominated Convergence theorem. Similarly, the second sum converges to zero provided that $hn^{(\xi-2)/\xi}= O(1)$ and $||\bs{b}|| < m_0 $, for some fixed constant $m_0 > 0$. 
The first part in the last term goes to zero with probability one by the uniform weak law of large numbers and the second part vanishes by direct calculation. 
We thus obtain
\beqn
\nonumber
\mathcal{L^\ast}(\bs{b}) &  =  &   \bs{b}^T \; W_n(x) -   2^{-1}  \bs{b}^T \Delta  \bs{b} \;  (1+ o_p(1)),
\eeqn
uniformly for $x \in \mathcal{X}$, where 
$$ W_n(x) =  \gamma_n \sum_{i=1}^n \ell_1 (\bar \eta(x,X_i) , U_{1i}, U_{2i}  )\; \bs{z}_{i,x} K((X_i -x)/h). $$ 
Using the quadratic approximation lemma \citep[p. 210]{Fan/Gijbels:1996}, 
 $$
 \hat{\bs{b}}   =  \Delta^{-1} \; W_n(u)   + o_p(1), 
$$
provided that $W_n$ is a stochastically bounded sequence of random vectors. The first entry of $ \hat{\bs{b}}$ directly yields the result, i.e.
\beqn
\nonumber
&&   \hspace{-0.2in}\gamma_n^{-1} \{ \hat \eta_h(x)- \eta(x) \}  =   \frac{ \gamma_n }{\sigma^2(x) f(x)} 
 \Bigg[   \sum_{i=1}^n  \ell_1(\eta(X_i),U_{1i},U_{2i})    K((X_i -x)/h)
 \\  \nonumber 
 &&  \hspace{-0.1in}+  \sum_{i=1}^n  \Big\{ \ell_1(\bar \eta(x, X_i),U_{1i},U_{2i})  -  \ell_1(\eta(X_i),U_{1i},U_{2i}) \Big\} K((X_i -x)/h) \Bigg] (1+ o_p(1)).
\eeqn\

\bigskip

\noindent {\bf Proof of Lemma \ref{lemma:2}}.
Note that
\beqn
\nonumber
T_{n1} & = &   \gamma_n^2 \sum_{k=1}^n    \frac{1}{  \sigma^2(X_k) f(X_k)  }    [\ell_1(\eta(X_k),U_{1k},U_{2k}) ]^2      \;  K\left(0\right)   \\ \nonumber
&&  \hspace{0.05in}+ \; \gamma_n^2 \sum_{k\neq i}   \frac{1}{  \sigma^2(X_k) f(X_k)}    \ell_1(\eta(X_i),U_{1i},U_{2i})  \ell_1(\eta(X_k),U_{1k},U_{2k})      \;  K((X_i - X_k)/h).
\eeqn
The approximation of the first term
\beqn
\nonumber
 \gamma_n^2 \sum_{k=1}^n    \frac{[\ell_1(\eta(X_k),U_{1k},U_{2k}) ]^2 }{  \sigma^2(X_k) f(X_k)  }         \;  K\left(0\right)
 =  h^{-1} K(0)  \E f^{-1}(X) + o_p(h^{-1/2}) 
\eeqn
yields the first result. We can decompose $T_{n2} = T_{n21} + T_{n22}$, where
\beqn \nonumber
  T_{n21} &= & \frac{1}{(nh)^2} \sum_{i=1}^n [\ell_1(\eta(X_i),U_{1i},U_{2i})]^2   \sum_{k=1}^n  \frac{ \ell_2(\eta(X_k),U_{1k},U_{2k}) }{  (\sigma^2(X_k) f(X_k))^2} K^2((X_i - X_k)/h),   \\ \nonumber
T_{n22}  &=& \frac{1}{n^2} \sum_{i\neq j} \ell_1(\eta(X_i),U_{1i},U_{2i}) \;\ell_1(\eta(X_j),U_{1j},U_{2j})  \Big\{ \sum_{k=1}^n  \frac{ \ell_2(\eta(X_k),U_{1k},U_{2k})}{  (\sigma^2(X_k) f(X_k))^2} 
\\ \nonumber
&& \hspace{3.2in}   \; K_h(X_i-X_k)    K_h(X_j-X_k) \Big\}.
\eeqn

\noindent We deal with $T_{n21}$ and $T_{n22}$ separately. For $T_{n21}$, note that
\beqn \nonumber
  T_{n21}  &=& \frac{1}{(nh)^2} \sum_{k=1}^n \ell_1(\eta(X_k),U_{1k},U_{2k})]^2  \frac{ \ell_2(\eta(X_k),U_{1k},U_{2k}) \;}{  (\sigma^2(X_k) f(X_k))^2} \; 
  K^2(0)   \\ \nonumber
  && \qquad +   \frac{1}{(nh)^2} \sum_{i \neq k} [\ell_1(\eta(X_i),U_{1i},U_{2i})]^2  \frac{ \ell_2(\eta(X_k),U_{1k},U_{2k}) \;}{  (\sigma^2(X_k) f(X_k))^2} \; 
 K^2((X_i - X_k)/h).
 \eeqn
The first sum can be shown to be
 \beq \nonumber
 \frac{1}{(nh)^2} \sum_{k=1}^n \sigma^2(X_k)  \frac{ \ell_2(\eta(X_k),U_{1k},U_{2k}) \;}{  (\sigma^2(X_k) f(X_k))^2}  K^2(0)    + o_p(h^{-1/2})  = O_p(n^{-1}h^{-2}).
 \eeq
Therefore, let
$$
V_n = \frac{2}{n(n-1)} \sum_{i < k} \Bigg[   \sigma^2(X_i)     \frac{ \ell_2(\eta(X_k),U_{1k},U_{2k}) }{  \{\sigma^2(X_k) f(X_k)\}^2} 
+    \sigma^2(X_k)     \frac{ \ell_2(\eta(X_i),U_{1i},U_{2i}) \;}{  \{ \sigma^2(X_i) f(X_i)\}^2}   \Bigg] 
K_h^2\big(X_k - X_i\big),
$$
and the second sum becomes
$(V_n+ o(1))/2   +   O_p\left(n^{-3/2}h^{-2}\right) + o_p(h^{-1/2})$.
The decomposition theorem for U-statistics \citep{Hoeffding:1948} allows us to show that 
$Var(V_n) =   O (n^{-1}h^{-2})$ as follows.
First note that the leading term of $V_n$ is  
$ - h^{-1} \E f^{-1}(X) \int K^2(t) dt.$
Hence, as $nh \to \infty$ and $h \to 0$, we obtain
\beq \nonumber
T_{n21} =  - h^{-1} \E f^{-1}(X) \int K^2(t) dt + o_p(h^{-1/2}).
 \eeq
Similarly, we can decompose $T_{n22}= T_{n221} + T_{n222}$ with
\begin{multline*}
T_{n221}  = \frac{2}{n}    \sum_{i < j}   \ell_1(\eta(X_i),U_{1i},U_{2i}) \;\ell_1(\eta(X_j),U_{1j},U_{2j})  \; \frac{1}{n} \Big\{ \sum_{k\neq i,j}  \frac{ \ell_2(\eta(X_k),U_{1k},U_{2k})}{  (\sigma^2(X_k) f(X_k))^2} 
\\
\; K_h(X_i-X_k)    K_h(X_j-X_k)    \Big\},
\end{multline*}
\begin{multline*}
T_{n222} =   \frac{K(0)}{n^2h} \sum_{i\neq j}  \ell_1(\eta(X_i),U_{1i},U_{2i}) \;\ell_1(\eta(X_j),U_{1j},U_{2j})  
\\
\times  \Big\{  \frac{ \ell_2(\eta(X_i),U_{1i},U_{2i})}{  (\sigma^2(X_i) f(X_i))^2} 
+    \frac{ \ell_2(\eta(X_j),U_{1j},U_{2j})}{  (\sigma^2(X_j) f(X_j))^2}  \Big\} K_h(X_i-X_j).
\end{multline*}
For $k \neq i, j$, define
\beq \nonumber
Q_{ijk, h}= \frac{ \ell_2(\eta(X_k),U_{1k},U_{2k})}{  (\sigma^2(X_k) f(X_k))^2} K_h(X_k-X_i)    K_h(X_k-X_j).
\eeq
It can be easily shown that 
$Var(n^{-1} \sum_{k\neq i,j} Q_{ijk, h})  =  O(n^{-1}h^{-2})$.
Then, 
$$
T_{n221} =   2n^{-2}(n-2)    \sum_{i < j}   \ell_1(\eta(X_i),U_{1i},U_{2i}) \ell_1(\eta(X_j),U_{1j},U_{2j})   
\E (Q_{ijk, h} | X_i,X_j)  + o_p(h^{-1/2}),
$$
where 
\beq \nonumber
 \E (Q_{ijk, h} | X_i,X_j) =  - \; \{h \;  \sigma^2(X_i) f(X_i)  \}^{-1} \int K(t)\; K((X_j- X_i)/h) dt.
\eeq
It is also easy to show
$Var(T_{n222}) = O \left(n^{-2}h^{-3}\right)$, implying
$T_{n222}=  o_p(h^{-1/2})$.

Combining $T_{n21}$, $T_{n221}$ and $T_{n222}$ yields
\bea
T_{n2}  =  - \;  \frac{1}{h} \E f^{-1}(X) \int K^2(t) dt 
-   \frac{2}{nh}    \sum_{i < j}    \frac{ \ell_1(\eta(X_i),U_{1i},U_{2i}) }  {   \sigma^2(X_i) f(X_i)  } \ell_1(\eta(X_j),U_{1j},U_{2j})    \\
\qquad 
\times   K \ast K(( X_j- X_i) /  h ) + o_p(h^{-1/2}). 
\eea \

\bigskip

\noindent {\bf Proof of Lemma \ref{lemma:3}}.
Recall that 
\beqn
\nonumber
W(n) &=& n^{-1} h^{1/2} \sum_{i \neq j}  \{\sigma^2(X_i) f(X_i)\}^{-2}    \ell_1(\eta(X_j),U_{1j},U_{2j})\; \ell_1(\eta(X_i),U_{1i},U_{2i}) \; \\ \nonumber
&&\qquad \qquad \qquad\qquad \qquad \qquad \{ 2 K_h(X_j- X_i) - K_h \ast K_h(X_j- X_i ) \}.
\eeqn
We shall show that $W_n$ satisfies conditions in Proposition 3.2.
Let
\beq \nonumber
W_{ij} = n^{-1}h^{1/2}  B_n(i,j)       \ell_1(\eta(X_i),U_{1i},U_{2i})\; \ell_1(\eta(X_j),U_{1j},U_{2j}),
\eeq
where
\beq  \nonumber
B_n(i,j) = b_1(i,j) +b_2(i,j) - b_3(i,j)-b_4(i,j),
\eeq
and
\beqn  \nonumber 
& b_1(i,j) = \displaystyle 2 K_h(X_j- X_i)\{\sigma^2(X_i) f(X_i)\}^{-2}    , \qquad \qquad   &  b_2(i,j) = b_1(j,i),
\\ \nonumber
&  b_3(i,j)  =  \displaystyle K_h \ast K_h(X_j- X_i ) \{\sigma^2(X_i) f(X_i)\}^{-2} ,   \qquad    & b_4(i,j) = b_3(j,i).
\eeqn

Thus we can write $W(n) = \displaystyle \sum_{i<j} W_{ij}$, and $W(n)$ is clean directly follows from the first Bartlett identity.
For the variance of $W(n)$, note that
$Var(W(n)) = \sum_{i <j} E(W_{ij}^2)$.
Thus we calculate $E[  \{  B_n(i,j)     \ell_1(\theta(X_i),U_{1i},U_{2i})\; \ell_1(\theta(X_j),U_{1j},U_{2j}) \}^2 ]$.
To simplify our presentation, let $\ell_{1i} = \ell_1(\theta(X_i),U_{1i},U_{2i})$ and denote the $m$-fold convolution at $t$  by $K(t,m) = K \ast \cdots \ast K(t)$.
Through direct calculations, we obtain
\bea   
&& \hspace{-0.33in}E(b_1^2(i,j)\;  \ell^2_{1i}  \; \ell^2_{1j}   ) =  E\left[  \frac{4}{h^2}   \frac{  \ell^2_{1i}  \; \ell^2_{1j} } {\{ \sigma^2(X_i) f(X_i)\}^2} K^2\left( \frac{X_j-X_i}{h} \right)   \right] 
\\ 
&&\hspace{0.3in}= \frac{4}{h^2} \int  \frac{ \sigma^2(X_1) } {\{ \sigma^2(X_1) f(X_1)\}^2}    \left\{   \int \sigma^2(X_2)    K^2\left( \frac{X_2-X_1}{h} \right)  f(X_2) dX_2 \right\}  f(X_1) dX_1
\\
&&\hspace{0.3in}= \frac{4}{h} \int \frac{f^{-2}(X_1)}{\sigma^2(X_1)}   \int \sigma^2(X_1) f(X_1) K^2(t) dt f(X_1) dX_1(1+O(h))
\\ 
&&\hspace{0.3in}= \frac{4}{h} K(0,2) Ef^{-1}(X)(1+O(h)).\eea
Similarly, 
\beqn
\nonumber
E(b_2^2(i,j)\;  \ell^2_{1i}  \; \ell^2_{1j}   )  & = &  4h^{-1} K(0,2) Ef^{-1}(X)(1+O(h)),  \\ \nonumber
E(b_3^2(i,j)\;  \ell^2_{1i}  \; \ell^2_{1j}   )  & = &  h^{-1} K(0,4) Ef^{-1}(X)(1+O(h)),  \\ \nonumber
E(b_4^2(i,j)\;  \ell^2_{1i}  \; \ell^2_{1j}   )  & = &  h^{-1} K(0,4) Ef^{-1}(X)(1+O(h)),  \\ \nonumber
E( b_1(i,j)      b_2(i,j)\; \ell^2_{1i}  \; \ell^2_{1j}  )      & = &  4 h^{-1} K(0,2) Ef^{-1}(X)(1+O(h)),  \\ \nonumber
E(   b_1(i,j)     b_3(i,j)   \; \ell^2_{1i}  \; \ell^2_{1j}   )  & = &  2 h^{-1} K(0,3) Ef^{-1}(X)(1+O(h)),  \\ \nonumber
E(   b_1(i,j)     b_4(i,j)   \;  \ell^2_{1i}  \; \ell^2_{1j}   )  & = &  2 h^{-1} K(0,3) Ef^{-1}(X)(1+O(h)),  \\ \nonumber
E(   b_2(i,j)     b_3(i,j)   \;  \ell^2_{1i}  \; \ell^2_{1j}  )  & = &  2 h^{-1} K(0,3) Ef^{-1}(X)(1+O(h)),  \\ \nonumber
E(   b_2(i,j)     b_4(i,j)   \; \ell^2_{1i}  \; \ell^2_{1j}   )  & = &  2 h^{-1} K(0,3) Ef^{-1}(X)(1+O(h)), \\ \nonumber
E(   b_3(i,j)     b_4(i,j)   \;  \ell^2_{1i}  \; \ell^2_{1j}    )  & = &  h^{-1} K(0,4) Ef^{-1}(X)(1+O(h)).
\eeqn
Thus, 
\beq
\nonumber
E[   B_n(i,j)  \ell^2_{1i}  \; \ell^2_{1j} ] =h^{-1} \{ 16 K(0,2) - 16 K(0,3) + 4K(0,4) \} Ef^{-1}(X)(1+O(h)).
\eeq
The leading term of $ \displaystyle n^{-2} h \sum_{i <j} E[  \{  B_n(i,j)   \ell^2_{1i}  \; \ell^2_{1j}  ]$ yields
\beq
\nonumber
\nu^\ast = 2 \{ 4 K(0,2) - 4 K(0,3) + K(0,4) \} Ef^{-1}(X) = 2 \;|| 2K - K\ast K||_2^2  \;Ef^{-1}(X).
\eeq
For the condition on $G_{I}$, note that
 $E(b_1(1,2) \ell_{11} \ell_{12}   )^4 =  E(b_3(1,2) \ell_{11} \ell_{12}   )^4 = O(h^{-3})$. 
Then $E(W_{12}^4 )=  n^{-4} h^2 O(h^{3}) $, which implies $G_I = O (n^{-2} h^{-1})= o(1)$. 
Similarly, the condition on $G_{II}$ can be verified by noting that 
$E(W_{12}^2 W_{13}^2)= O(E(W_{12}^4 )) = O  (n^{-4} h^{-1}). $
Thus, $G_{II} = O(n^{-1} h^{-1})= o(1)$. For the last condition we need to check the order of $E(W_{12}W_{23} W_{34}W_{41})$. 
Calculations for few terms yield,
\beqn
\nonumber
E(b_1^2(1,2)  b_1^2(2,3) b_1^2(3,4) b_1^2(4,1)    \;  \ell'^2_1  \; \ell'^2_2 \;  \ell'^2_3  \; \ell'^2_4  )  & = &  O(h^{-1})  \\ \nonumber
E(b_1^2(1,2)  b_1^2(2,3) b_1^2(3,4) b_3^2(4,1)    \;  \ell'^2_1  \; \ell'^2_2 \;  \ell'^2_3  \; \ell'^2_4  )  & = &  O(h^{-1})  \\ \nonumber
E(b_1^2(1,2)  b_1^2(2,3) b_3^2(3,4) b_3^2(4,1)    \;  \ell'^2_1  \; \ell'^2_2 \;  \ell'^2_3  \; \ell'^2_4  )  & = &  O(h^{-1})  \\ \nonumber
\nonumber
E(b_1^2(1,2)  b_3^2(2,3) b_3^2(3,4) b_3^2(4,1)    \;  \ell'^2_1  \; \ell'^2_2 \;  \ell'^2_3  \; \ell'^2_4  )  & = &  O(h^{-1})  \\ \nonumber
E(b_3^2(1,2)  b_3^2(2,3) b_3^2(3,4) b_3^2(4,1)    \;  \ell'^2_1  \; \ell'^2_2 \;  \ell'^2_3  \; \ell'^2_4  )  & = &  O(h^{-1}). 
\eeqn
Since terms with other combinations will be of the same order, we conclude that 
$$E(W_{12}W_{23} W_{34}W_{41}) =  n^{-4} h^2 O(h^{-1}) =O(n^{-4} h), $$
and $G_{IV} = O(h)= o(1)$. This completes the proof.

\end{document}